\documentclass[journal]{IEEEtran}
\usepackage{amsmath,amsfonts}
\usepackage{algorithmic}
\usepackage{textcomp}
\usepackage{stfloats}
\usepackage{url}
\usepackage{verbatim}
\usepackage{graphicx}
\usepackage{cite}

\begin{document}

\title{Experimental Investigation of 5G Base Station functionalities in Reverberation Chamber at Millimeter-Wave}

\graphicspath{{./}}

\title{Experimental Investigation of 5G Base Station functionalities in Reverberation Chamber at Millimeter-Wave}
\author{Michele Colombo, Riccardo Diamanti, Luca Bastianelli, Gabriele Gradoni, Emanuel Colella, Valter Mariani Primiani, Franco Moglie, Davide Micheli
\thanks{Michele Colombo is with Nokia Networks Italia, 20871 Vimercate, Italy\\
Riccardo Diamanti is with TIM S.p.A., Via Guido Miglioli 11, 60131 Ancona, Italy\\
Luca Bastianelli, Emanuel Colella, Valter Mariani Primiani and Franco Moglie are with Dipartimento di Ingegneria dell'Informazione, Universit\`a Politecnica delle Marche, 60131, Ancona, Italy and with Consorzio Nazionale Interuniversitario per le Telecomunicazioni (CNIT), 43124 Parma, Italy\\
Gabriele Gradoni is with Institute for Communication Systems (ICS), University of Surrey, GU2 7XH Guildford, U.K. and with Department of Computer Science and Technology, University of Cambridge, CB3 0FD Cambridge, U.K.\\
Davide Micheli is with TIM S.p.A., 00189 Rome, Italy\\
``This work has been supported by EU H2020 RISE-6G project under the grant number 101017011.''}
}

\maketitle

\begin{abstract}

The performance and functionalities of a commercial fifth generation base station are evaluated inside the reverberation chamber at the mmWave frequency range.
The base station capability to operates in different propagation environment conditions reproduced by the reverberation chamber is investigated. 
Throughput, modulation code scheme and beamforming are analyzed for different real life scenarios both in uplink and downlink.
Experimental results inform network operators in their evaluation of the base station operation: i) in many scenarios within a laboratory; ii) in the assessment of whether expected benefit justifies the additional costs in an operating actual network.

\end{abstract}

\begin{IEEEkeywords}
5G, base station, mmWave, reverberation chamber, key-performance indicators, live
radio access network, throughput, RSRP, beamforming.
\end{IEEEkeywords}

\section{Introduction}
\label{sec:introduction}
Mobile communication has rapidly increased over last years, evolving from only voice services to dense interconnected
environment serving multiple services. Those services run on a system based on the same infrastructure that is able to support a large number of applications. The system provides connection among a massive number of devices ensuring low latency and high-speed communication, such as autonomous vehicles scenarios, smart cities and so on.
In the next few years, it is forecast that the data traffic and services will enormously increase, the 5G represents not only a solution but the revolution we need.
The 5G wireless technology is a significant advancement over previous generations. It solves all prior disadvantages, such as a lack of coverage,
lack of performance at cell edges, and dropped calls.
In wireless applications the propagation environment may arise some problems in the overall system operation.
Some techniques can be adopted in the 5G system in order to mitigate and improve the connectivity.
In particular, BSs will be equipped by massive MIMO antennas which provide eMBB
to be leveraged for many applications.
The 5G technology introduces novelty and its application will revolutionize many areas, such as smart city, IoT,
M2M communication, health care and we can go on for much longer listing the most varied applications~\cite{ITU_vision, 3GPP_market_technology, 2019_5G_evolution}.
In the not-too-distant future, as early as five years from now, it is expected a huge increment of mobile traffic~\cite{sakaguchi2015millimeter, 2017_Vincenzo, report_ericsson}.
Mobile operators are in a race to deploy technologies in order to support networks with a massive number of devices, interconnected each others able leverage high bandwidth in
particular at the mmWave frequency range, such as $1$~Gbps or more.
On the one hand mobile operators have the possibility to virtualize and slicing network in order to support different product/services.
On the other hand, wireless networks can be integrated with passive/active devices able to improve the radio communication and they enable functionalities taking advantage of
mobile edge computing. Those devices are RISs which can be adopted for smart and programmable radio environments, thus by giving new opportunities and new paradigms~\cite{strinati20216g, 2021_rise6g_perspective, 2022_EuCNC}.
From the mobile operator point of view, the perspective of reducing: the transmitting power, interference, electromagnetic field exposure~\cite{9103530, 9690179} to non-intended user and so on, makes the technology very appealing.
Beyond the current frequency bands, mobile operators are going to occupy the mmWave frequency bands~\cite{2017_where_when_how_Strinati, 2015mmWave_overview} 
that guarantee a part of spectrum which is not fully licensed.
This spectrum allows to support a high data rate and large number of interconnections.
On the contrary, mmWave band exhibits a large path loss, in particular when the signal encounters an obstacle.
To overcome this issue mobile operator has different options such as the adoption of smart antennas, a mix of transmission techniques or either the option to be discarded i.e. the increase in transmitted power.
In order to perform measurements with a commercial Nokia 5G BS, equipped by a smart antenna with beamforming, we take advantage of the characteristics of the RC~\cite{corona_1980, KildalProcIEEE2012}.
The RC can be efficiently adopted to carry out measurements of total radiated power of general electronic equipment~\cite{iec-61000-4-21}
and recently applied to base station radiated power determination~\cite{9618469}.
It is interesting to note that the RC methodology is
applicable to the higher 5G frequencies too, which are $28$~and~$39$~GHz and regardless of beamforming technology~\cite{2022_Gifuni_power}.
The method has been also investigated in terms of associated uncertainty~\cite{9676572} affected by the coherence bandwidth
due to the large bandwidth adopted by modern communication systems.
The availability of a very rich multipath environment like an RC is useful to carry out complete OTA on wireless systems~\cite{holloway_rician_2006}
because real life propagation conditions can be reproduced.
The RC is an electrically large cavity where the electromagnetic field is statistically uniform, isotropic and randomly
polarized within a defined region, called working volume~\cite{corona_1996, hill-plane_wave}.
The whole surface of the RC is covered by metal. The RC is equipped by stirrers that continuously rotate in order to mix the
electromagnetic field and thus to replicate dynamic propagation conditions~\cite{RC_2011_Remley}.

The RC is a richly multipath environment that does not represent the majority of real life scenarios, but by adding absorbing materials
the RC load increases and the corresponding time delay spread progressively decreases thus allowing us to use it for  wireless tests.
The presence of absorbing panels reduces reflected signals and we are able to control the multipath richness in order to emulate the desired real
life propagation environments~\cite{RC_2012_Holloway, 2015EMCS_Bastianelli_PDP,2021wang}.
By properly setting the RC loading, time delay spread reaches typically values encountered in real life environments.
In particular, to this purpose we can consider few parameters, such as the PDP and time delay spread $(\tau)$, the $Q$-factor and the Rician $K$-factor.
Those values are tuned by varying the number of absorbing panels, planar and/or pyramidal.
The time delay spread is an indication of the spread affecting the signal components,
as they are coming to the receiver.
The decay time denotes the time after which the reflected component is negligible and it directly influences the coherence time.
Finally, the $Q$-factor is the capability to maintain the transmitted energy inside the RC, whereas the $K$-factor is the ratio
between the direct and scattered energy~\cite{2015_chen_Kfactor, holloway_rician_2006, lemoine2011mode}.
The RC environment has been already used to test the 4G-LTE
system~\cite{2015_Micheli_LTE_CA, RC_2015_Micheli, RC_2017_LTE_uplink, 2018_Micheli_high_speed,2017_LTE_testing,2021_MICHELI_MIMO,2020_Barazzetta_VoLTE, 2015EMCS_Bastianelli_PDP, 2016_barazzetta_EuMC}.

In this paper, we present results of a measurements campaign executed in the RC of the Universit\`a Politecnica
delle Marche, in cooperation with Telecom Italia and Nokia, focused on the mmWave frequency range.

\section{Set-up description}

\begin{figure}[!t]
\centering
\includegraphics[width=2.2in]{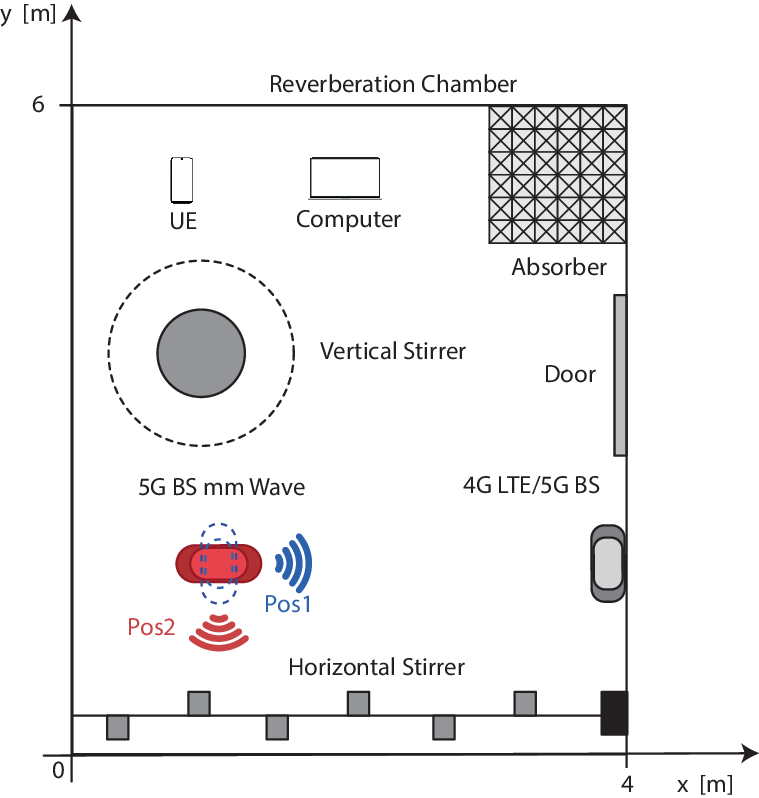}
\caption{Schematic description of the set-up. During measurements we considered two antenna orientations labeled as ``Pos1'' and ``Pos2'' respectively.}
\label{fig:setup1}
\end{figure}

The RC used in this measurement campaign has dimensions 4~m wide ($\hat{x}$), 6~m long ($\hat{y}$) and 2.5~m high ($\hat{z}$). It is large enough to hold antennas, UEs, absorbers
and all instruments we need.
The RC is equipped by two stirrers, a vertical Z-folded stirrer and a horizontal stirrer.
Figure~\ref{fig:setup1} shows the 2D schematic of the adopted set-up.
During measurements, the vertical stirrer always remained stationary, while the horizontal stirrer rotated in continuous way in some tests.
During tests we changed the antenna orientation: i) in ``Pos1'' it faces towards the RC door (highlighted in blue in \figurename~\ref{fig:setup1}); ii) in ``Pos2'' it faces towards the RC horizontal stirrer (highlighted in red in \figurename~\ref{fig:setup1}).

\begin{figure}[!t]
\centering
\includegraphics[width=1.5in]{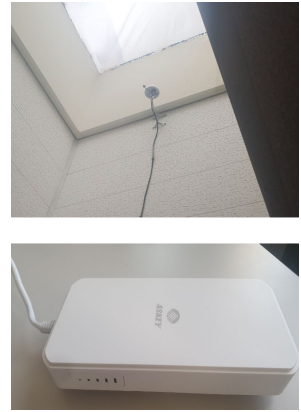}
\caption{The top picture shows the GPS antenna used to synchronize the 5G pilot signal. The bottom picture reports the Askey UE used to collect data.}
\label{fig:UE}
\end{figure}

In this measurement campaign we used a 5G Nokia BS, the Askey terminal and the QXDM software in order to collect data from the user point of view. Figure~\ref{fig:UE} shows the GPS antenna installed in our laboratory and the UE Askey terminal.
From the mobile operator point of view, Nokia operator is able to have direct access to BS data thanks to the BS manufacturer proprietary software.
The radio modules, both 4G LTE and 5G, were connected to the TIM's live core network with an optical fiber.
In this work two real life scenarios were emulated, more precisely a commercial (\#1) and a residential (\#2) ones.
The time delay spread effectively quantifies the environment multipath propagation
condition, as specified by
ITU~\cite{ITU_outdoor, ITU_indoor}.
In order to characterize the chamber we collected the $S_{21}$ parameters by a VNA between two horn antennas.
The investigated frequency range during these tests is from to $27$ to $29$~GHz. 
By performing an IFT we evaluated the chamber impulse response from
\begin{equation}
h \left ( t \right ) = \textrm{IFT} \left [ S_{21} \right ] \, .
\label{eq1}
\end{equation}
Then, the PDP is evaluated by 
\begin{equation}
\textrm{PDP} \left ( t \right )=\left \langle \left | h \left ( t \right ) \right |^2 \right \rangle _N \, ,
\label{eq2}
\end{equation}
where $\left < \cdot \right >$ denotes the ensemble average over the $N$ stirrer positions. When the PDP is computed, a threshold were considered.
This value defines the minimum signal level considered to compute the $\tau_{RMS}$, we set the threshold equal to $-20$~dB~\cite{RC_2010_Genender, ITU_6to100GHz}.
The time delay spread $(\tau)$ and its root mean square $(\tau_{RMS})$ are related to
the PDP and computed by eq.~\eqref{eq:tau} and eq.~\eqref{eq:tau_ave} respectively.
\begin{equation}
\tau_{ \textrm{RMS}}= \frac{\sqrt{\int_0^\infty {\left ( t - \tau_{ave} \right )^2 \textrm{PDP} \left ( t \right )dt}}} {\int_0^\infty { \textrm{PDP} \left ( t \right )dt}} \, ,
\label{eq:tau}
\end{equation}
\begin{equation}
\tau_{ave} =\frac{ \int_0^\infty {t \textrm{PDP} \left (t\right ) dt}}{\int_0^\infty { \textrm{PDP} \left (t\right ) dt}} \, .
\label{eq:tau_ave}
\end{equation}

\begin{figure}[!h]
\centering
\includegraphics[width=2.8in]{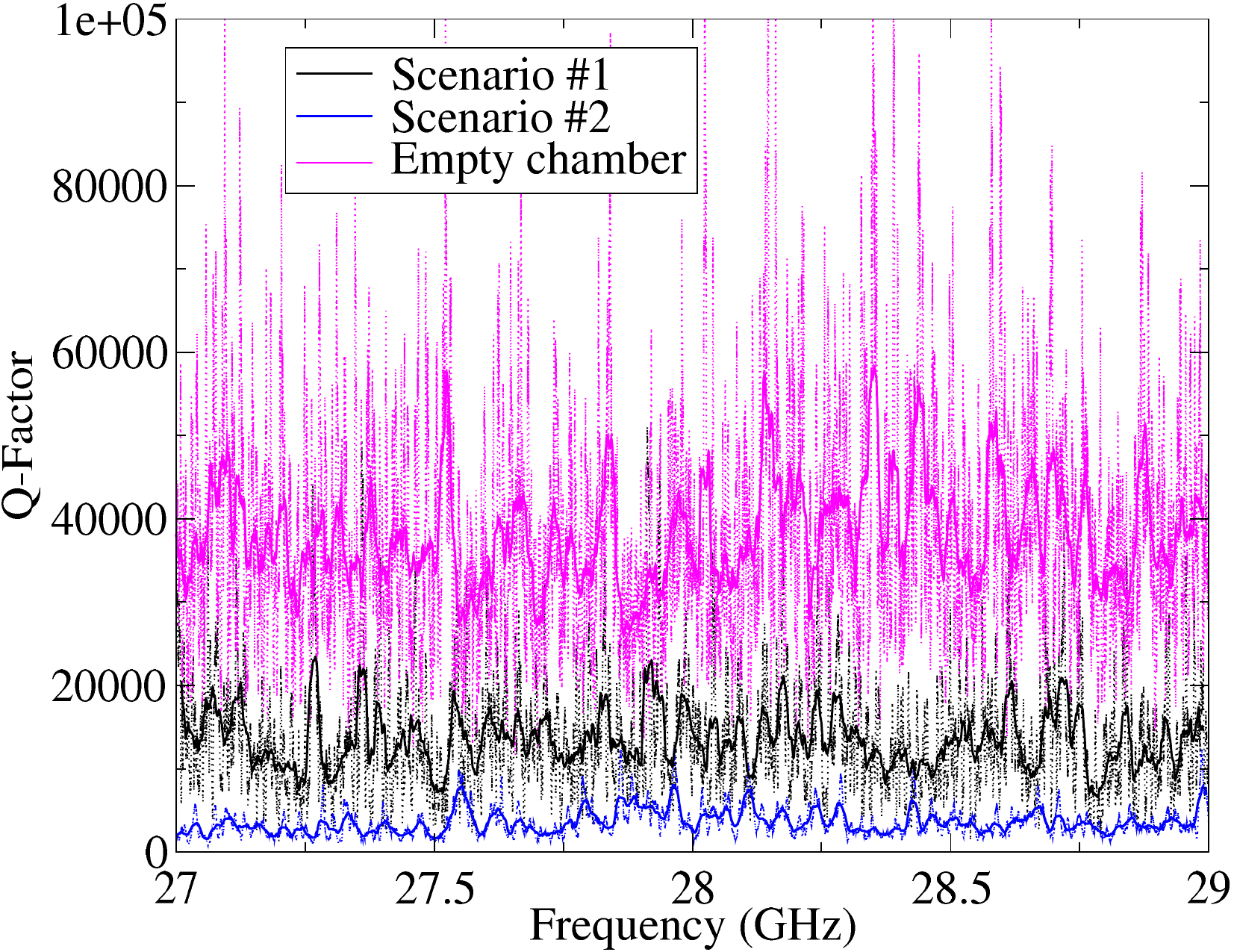}
\caption{Quality factor for the different chamber loading condition, EC, ML and UHL. Dotted lines denotes raw data whereas continuous lines are the raw data are after by applying a sliding average in a window of 400 frequency points.}
\label{fig:factor}
\end{figure}

\begin{figure}[!t]
\centering
\includegraphics[width=2.8in]{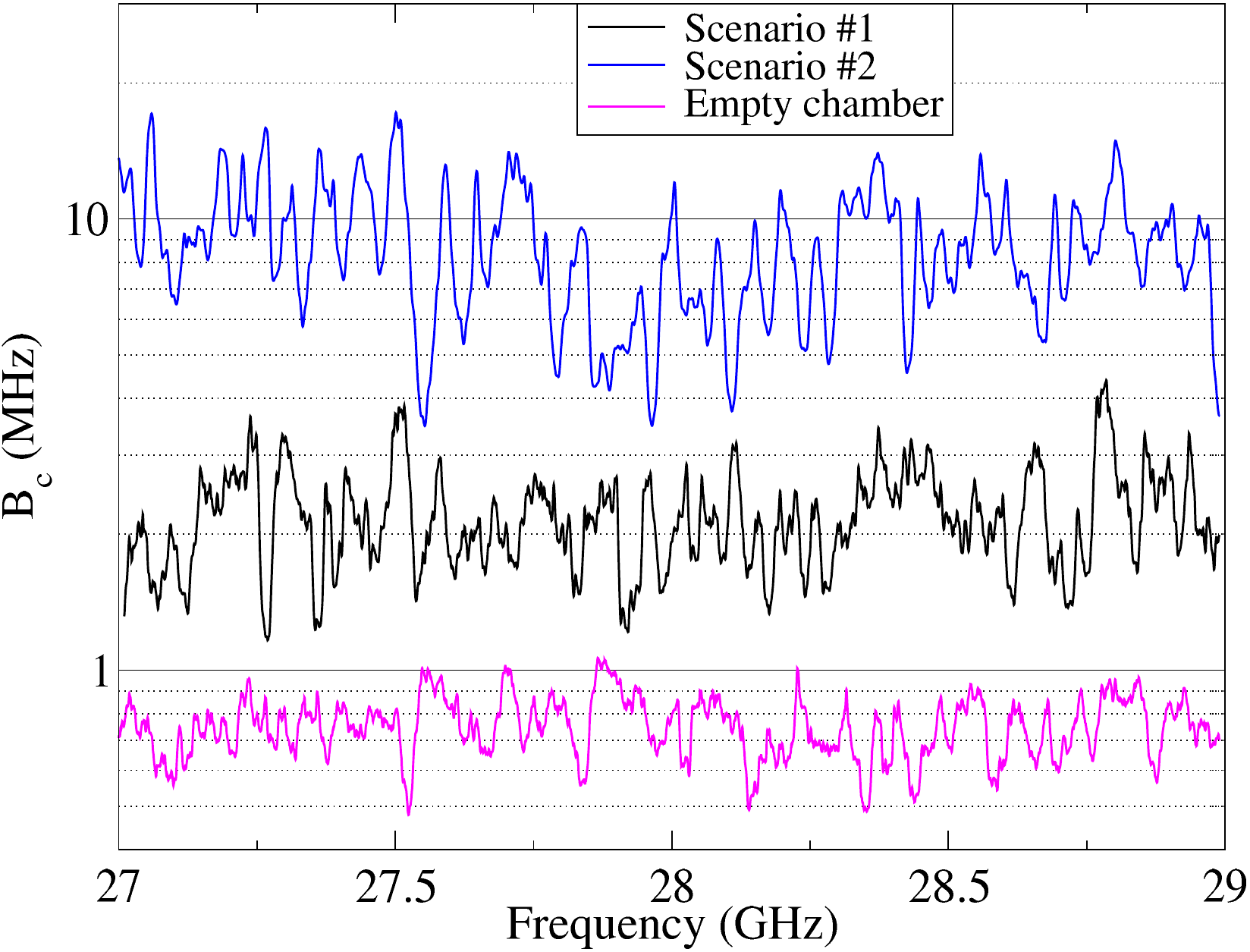}
\caption{Coherence bandwidth.}
\label{fig:bc_plot}
\end{figure}

The $Q$-factor of the chamber conveys the ratio between the stored energy and the dissipated power within the RC and it is computed by
\begin{equation}
Q = \frac{16 \pi^2 V \left < \left | S_{21} \right | ^2 \right > _N}{ \eta_{TX} \eta_{RX} \lambda ^3 } \, ,
\label{eq:eq3}
\end{equation}
where $V$ is the volume of the RC, $\lambda$ is the free-space wavelength, $S_{21}$ are the scattering parameters
and $\eta_{TX}, \eta_{RX}$ are the transmitting and receiving antenna total
efficiencies, respectively. By means of the $Q$-factor it is possible to evaluate the decay time of the chamber by~\cite{iec-61000-4-21}:
\begin{equation}
\tau = \frac{Q}{\omega} \, ,
\label{eq6}
\end{equation}
where $\omega$ denotes the angular frequency. The $B_c$ is computed by the relationship:
\begin{equation}
B_c = \frac{f}{Q} \, ,
\label{eq4}
\end{equation}
where $f$ denotes the frequency. 
Figure~\ref{fig:factor} shows the $Q$-factor of the explored scenarios.
The chamber $Q$-factor rapidly reduces when absorbing materials is inserted.
The final effect expected passing from ``Scenario \#1'' to ``Scenario \#2'' is a reduction in the energy inside the RC, giving the same transmitted power.
Consequently, the signal level experienced by the UE will reduce. Figure~\ref{fig:bc_plot} reports the coherence bandwidth. The $B_c$ essentially is the band within which the propagation properties of the channel do no vary appreciably.


\begin{figure}[!t]
\centering
\includegraphics[width=2.3in]{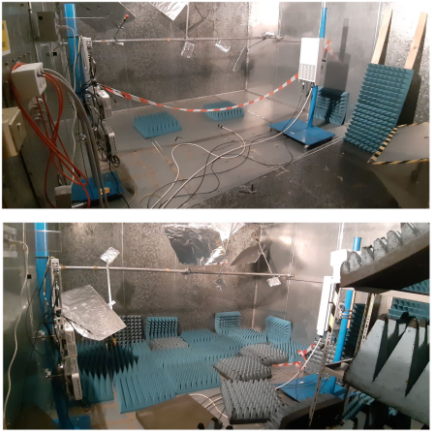}
\caption{Inner view of the RC. On the top the ML condition whereas on the bottom the UHL condition. Moreover, for the antenna orientation the top picture denotes ``Pos1'' whereas the bottom picture refers to ``Pos2''.}
\label{fig:low_load}
\end{figure}

Explored chamber loading configurations are:
\begin{itemize}
\item EC: within the chamber there are only cables, PC, UE and the BS. In this case, the $Q$-factor is high, losses mechanism are
offset by the high frequency range;
\item ML chamber (``Scenario \#1''): nine anechoic absorbing panels (Emerson \& Cuming VHP-8-NRL) are added within the RC. The setup is reported in the top picture of \figurename~\ref{fig:low_load}. The $Q$-factor
decreases w.r.t. the EC case as showed in \figurename~\ref{fig:factor}. This load condition emulates a commercial scenario;
\item UHL chamber (``Scenario \#2''): in this scenario forty-five anechoic absorbing panels (Emerson \& Cuming VHP-8-NRL),
four anechoic absorbing panels (Emerson \& Cuming VHP-18-NRL) and seven planar
absorbing panels (Emerson \& Cuming ANW-77) were placed on the RC floor and resting on wooden boards lean in the walls. The setup is reported in the bottom picture of \figurename~\ref{fig:low_load}. In this situation the $Q$-factor
decreases w.r.t the ML condition. This load condition emulates a residential scenario.
\end{itemize}

The more the chamber is loaded, the more the $Q$~factor is reduced.
Combinations of number and position of absorbers within the RC differently affect the overall loading behavior~\cite{2015EMCS_Bastianelli_PDP}.
In this measurement campaign we arranged absorbing panels on the floor, on walls of the chamber and in a metallic organizer in order to control: i) the $Q$-factor, ii) the PDP and iii) $\tau_{RMS}$~\cite{Holloway_PDP,Genender_PDP,Fielitz}.
We performed measurements at the mmWave frequency range, in this scenario the transmitted beam has a high directivity and this implies a narrow beam~\cite{Balanis-2012-antenna}.
The RC exhibits a different behavior at mmWave w.r.t. sub-6 GHz analyzed in the past~\cite{2015_Micheli_LTE_CA, RC_2015_Micheli, RC_2017_LTE_uplink, 2018_Micheli_high_speed,2017_LTE_testing,2021_MICHELI_MIMO,2020_Barazzetta_VoLTE, 2015EMCS_Bastianelli_PDP, 2016_barazzetta_EuMC}.
In this way, unless you want to block the direct connection, the position of the absorbing material is not so relevant in a multipath also considering that the signal after some reflections fades considerably.
At the mmWave range the PDP fades more quickly than lower frequencies, i.e. sub-6~GHz~\cite{2015EMCS_Bastianelli_PDP}, and that is precisely why new 5G systems use beamforming and high gain antennas that compensate the channel in hostile situations while maintaining a radio link acceptable.
As we will see shortly, the 5G active smart antenna is able to select the best configuration, in terms of selected beam(s), to create and maintain the mobile connection.

\begin{figure}[!h]
\centering
\includegraphics[width=2.8in]{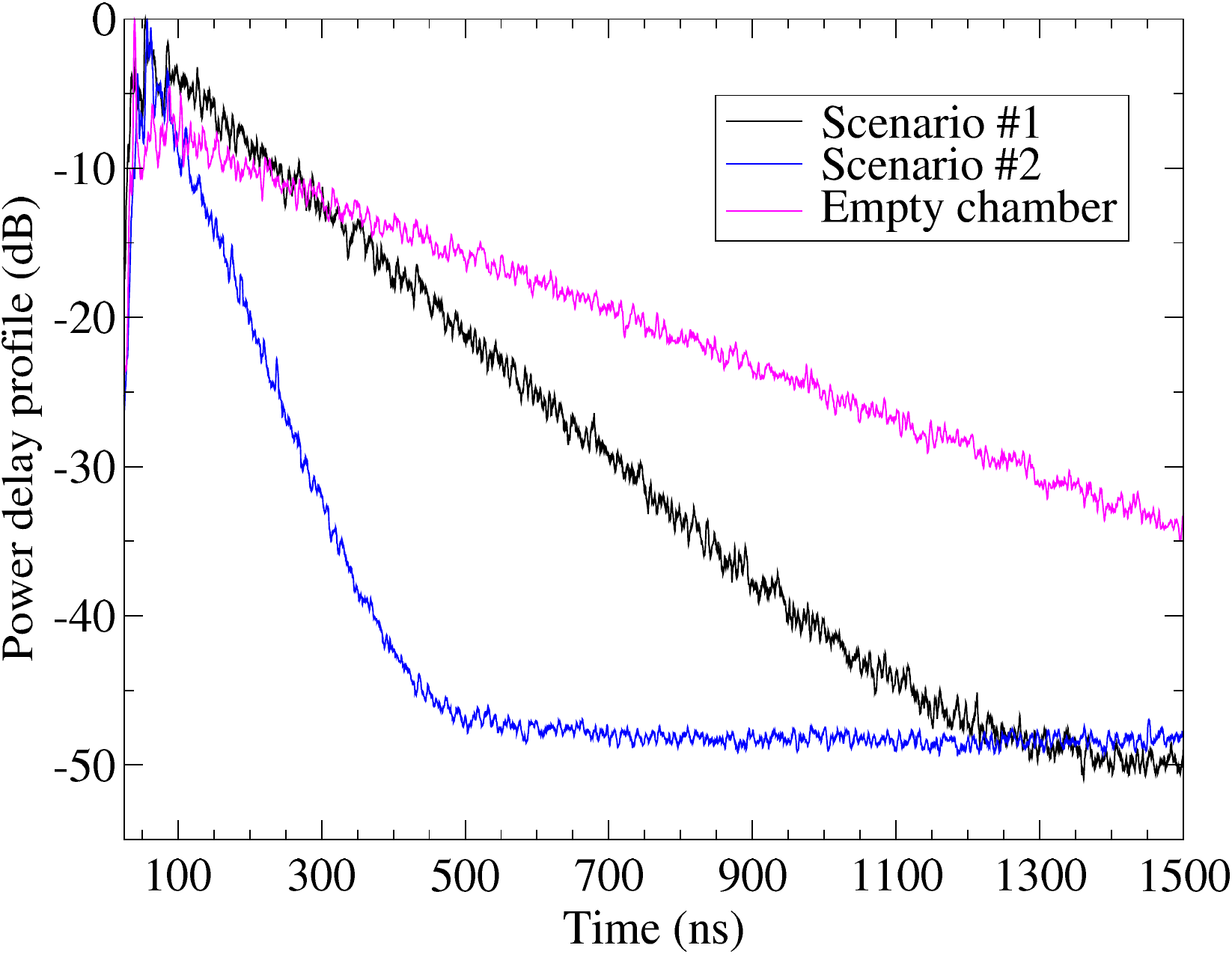}
\caption{Power delay profile for different chamber loading condition.}
\label{fig:PDP}
\end{figure}

In \figurename~\ref{fig:low_load} are reported the ML and UHL setup used during tests.
In order to emulate a real-life scenario, the number of absorbing materials was progressively increased until
the computed $\tau_{RMS}$ matched its value reported in the standard~\cite{ITU_outdoor, ITU_indoor}.
The PDP of those scenarios are reported in \figurename~\ref{fig:PDP}. In the UHL scenario the PDP exhibits a greater reduction w.r.t ML case.
The PDP variation affects the $\tau_{RMS}$, the computed values are reported in Table~\ref{tab:tds} considering a threshold of $-20$~dB~\cite{RC_2012_Holloway}.
The PDP reduction occurs for a larger interception of rays by absorbers before they reach the receiving antenna.

\begin{table}
\caption{Time delay spread applying a threshold of -20~dB.}
\label{tab:tds}
 \begin{tabular}{c|l|c}
  Frequency (GHz) & Environment &  $\tau_{RMS}$ (ns) \\ \hline
  28  & \#1: Commercial   &   65 \\ 
  28  & \#2: Residential  &   19 \\ 
 \end{tabular}
\end{table}
In Table~\ref{tab:tds} are reported the computed $\tau_{RMS}$, its reduction occurs for a larger interception of rays by absorbers before
they reach the receiving antenna.

\section{BS operating conditions}

\begin{figure}[!ht]
\centering
\includegraphics[width=1.6in]{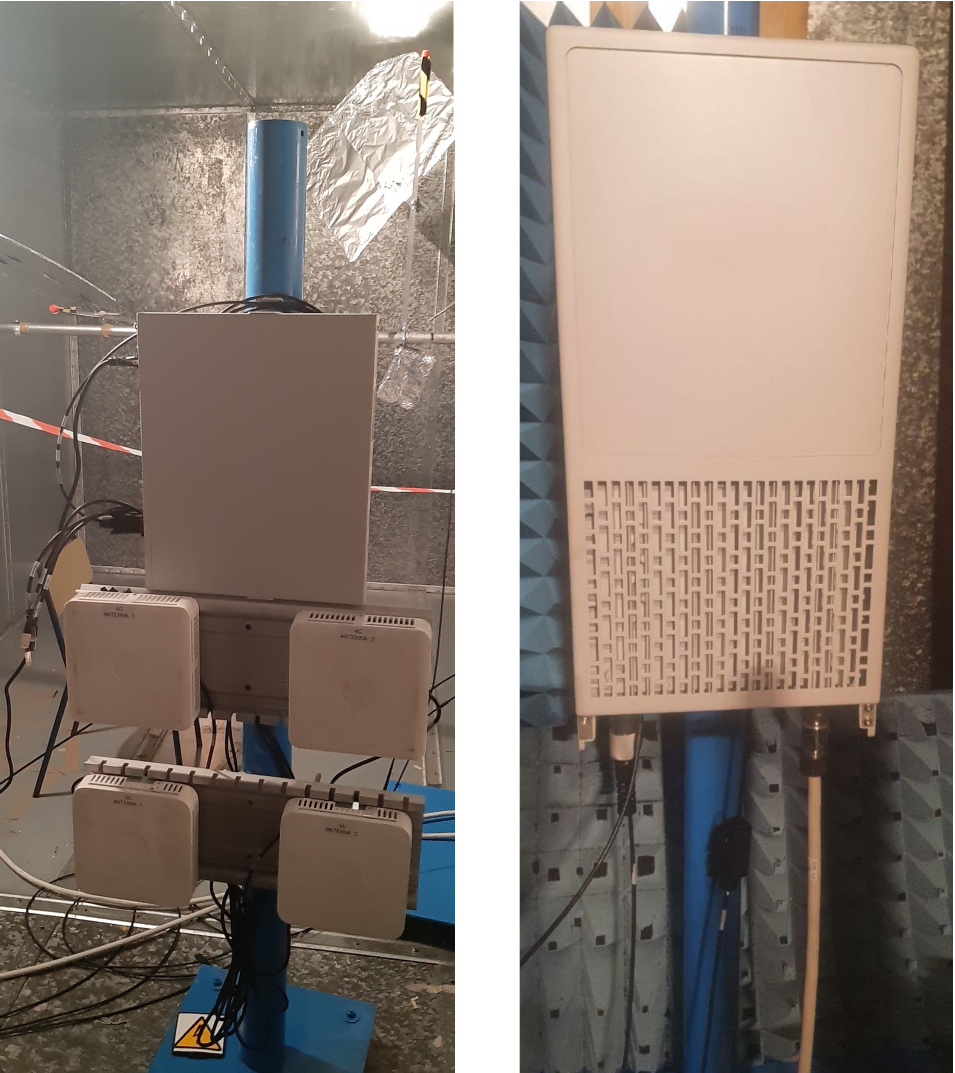}
\caption{Pictures of the BS antennas. On the left, on the top the 5G antennas at 3.7 GHz whereas on the bottom the 4G LTE RF modules. On the right the 5G mmWave antenna with analog beamforming.}
\label{fig:BSs}
\end{figure}

\begin{figure}[!ht]
\centering
\includegraphics[width=1.8in]{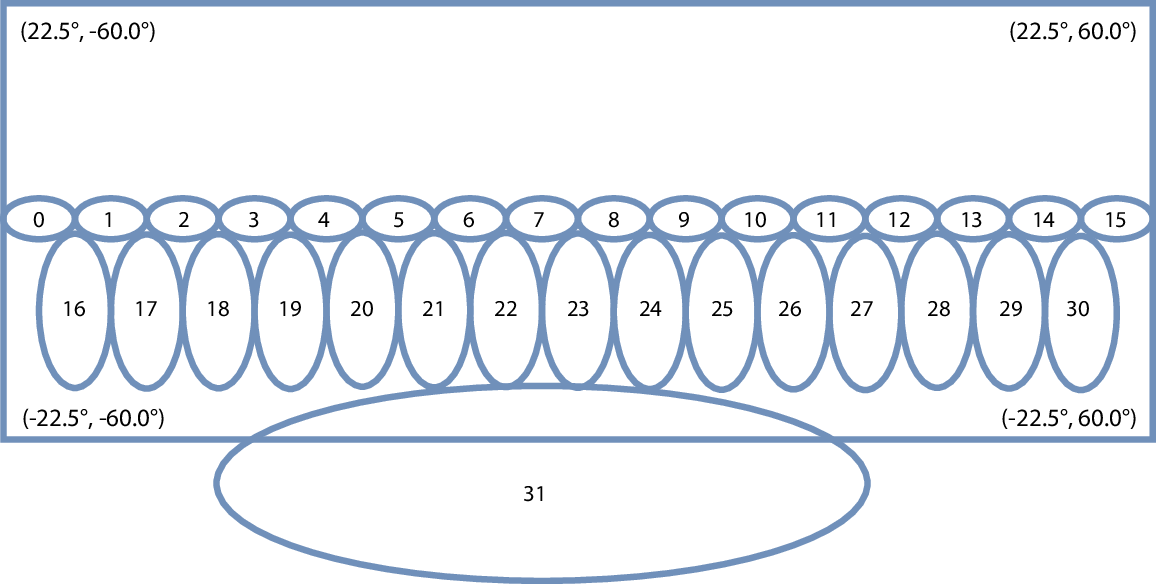}
\caption{Selected pattern configuration of the 32 beams.}
\label{fig:beam}
\end{figure}

The AEUB Nokia active antenna architecture used in these measurements is no stand alone, it needs the control signal by the 4G LTE modules.
Figure~\ref{fig:BSs} shows the 5G and 4G LTE RF antenna modules. The mmWave antenna, the right one showed in~\figurename~\ref{fig:BSs} has dimensions $603$~mm height, $304$~mm width and $124$~mm thick.
In spite of the transmitted power due to the control signal that lean on the LTE, it exhibits a little impact in the UL scenario
and it is negligible on the total transmitted power.
The mmWave 5G antenna is configured by MIMO~$2\times2$, with a bandwidth of $1$~GHz and a maximum power of $28$~dBm for each transmitter
whereas the antenna gain is $28$~dBi.  The total average EIRP is $60$~dBm.
The supported operating frequency band are from~$27.50$~to~$28.35$~GHz and from~$26.50$~to~$29.50$~GHz. The operating center frequency in our tests was~$26.95$~GHz.
The single SSB has maximum power of $16$~dBm whereas the transmitted power of the UE is $23$~dBm.
During tests we enabled the carrier aggregation, in spite of we used only $1$ cell.
The MIMO~$2\times2$ is present both in DL and UL.
In these tests, the maximum attainable throughput is $550$~Mbps for the DL and $140$~Mbps in UL. 
The frame for both beamforming and broadcast is composed by $40$ slots, each with a duration of $0.125$~ms.
In fact, every $0.5$~ms the system sends $1$~SSB and the terminal highlight which beam exhibits the best RSRP, then move on the throughput.
The AEUB Nokia has an analog beamforming, no other techniques are implemented such as refining beam.
The mobile operator can choose from multiple beam set configurations each of them has a different beam pattern.
For preliminary tests we fix the $32$~beams configuration reported in \figurename~\ref{fig:beam}, with one centered beam and the others on vertical panels.
In this configuration, the beam set azimuth opening angle is from $-60$~deg to $+60$~deg whereas the elevation opening angle is from $+87$~deg to $+123$~deg.
\begin{figure*}[h]
\centering
\includegraphics[width=4.1in]{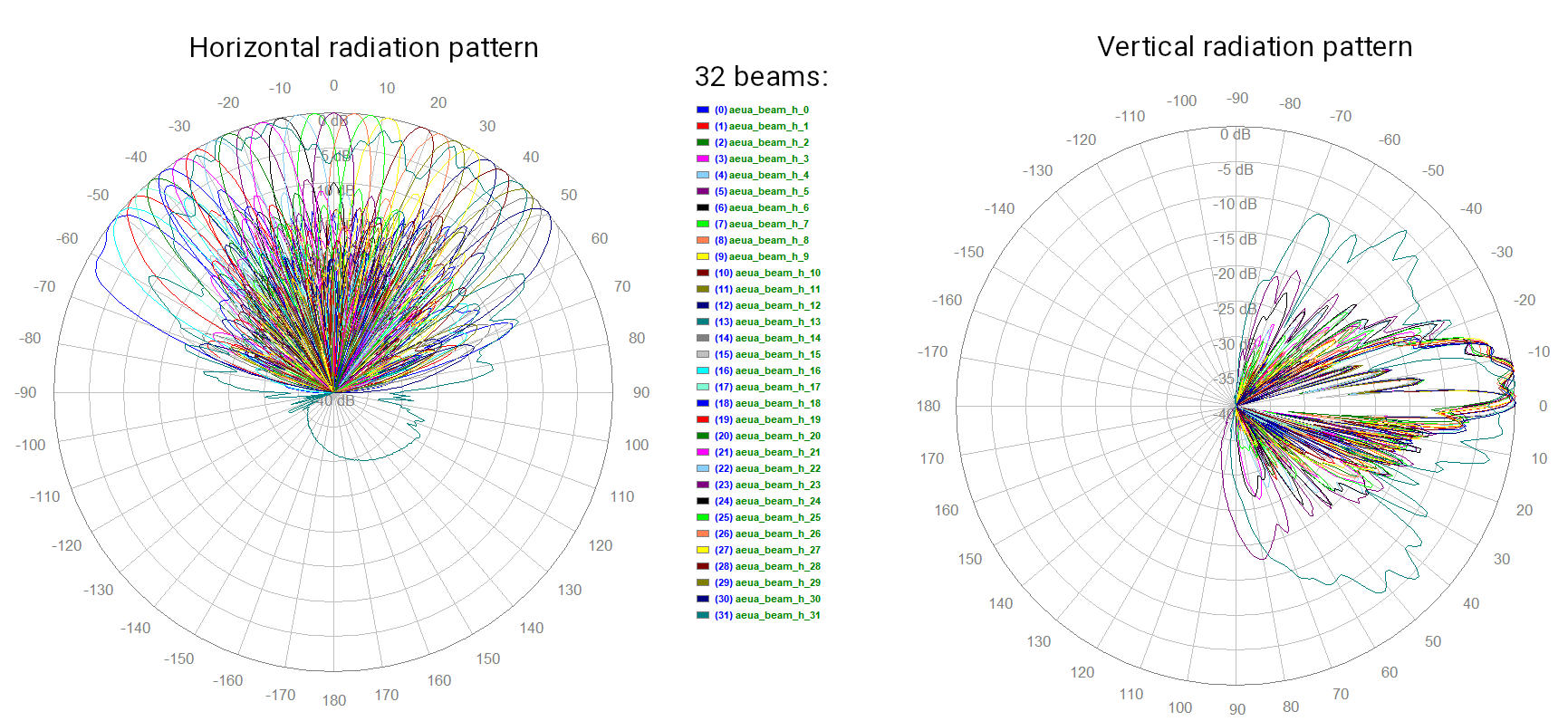}
\caption{Radiation pattern of the 32 beams. The envelop of of beams can cover from -60~deg to +60~deg azimuth angle.}
\label{fig:beam_overall}
\end{figure*}

The envelope of all beams is showed in \figurename~\ref{fig:beam_overall}, in accordance with what is shown in \figurename~\ref{fig:beam}. The radiation patterns are plotted by using a proprietary software provided by the BS vendor.
It is also possible to choose and set only one beam for tests.
It is worth noticing that in a real situation, if a beam points to an obstacle that intercepts the chosen beam, the antenna switches to the close beam.
The distance between maximum and zeros of the radiation pattern is very small.
In this context the overall gain remains high in order to maintain a high RSRP.
This fact is opposite to the LTE scenario where the relocation of the receiver can cause an evident degradation in terms of RSRP.

The UE used to monitor the evaluation of the BS is an Askey RTL 6305, which has MIMO~$2\times2$ in UL.

\section{Results}
In all tests, in every radio configuration (also when -60 dBm SSB), MIMO~$2\times2$  were always maintained. 
It can be explained by the paths uncorrelation on the tested environments.
The throughput reduction does not depend on the diversity scheme but it is due to the SNR degradation, both from DL and UL.
\begin{figure}[!h]
\centering
\includegraphics[width=1.2in]{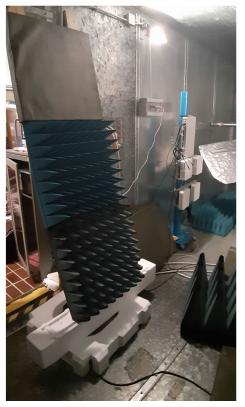}
\caption{Door of the RC covered by two VHP-18-NRL and one ANW-77 absorber.}
\label{fig:portacoperta}
\end{figure}
\begin{figure}[!h]
\centering
\includegraphics[width=2.1in]{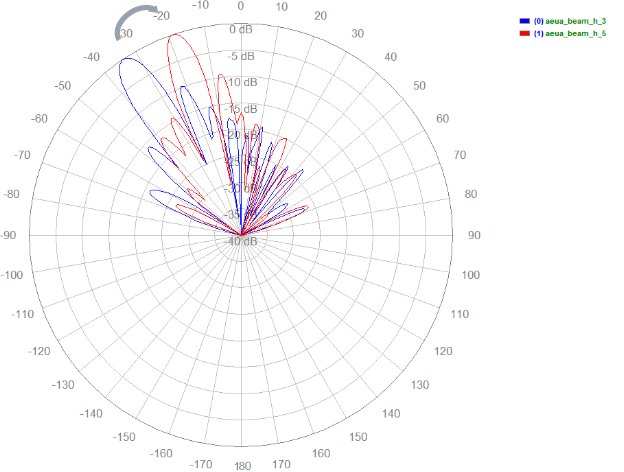}
\caption{Horizontal radiation pattern when the door is blocked by absorbing panels in the ``Pos1'' case. The BS switches from beam\#3 (blue) to beam\#5 (red).}
\label{fig:beamtest}
\end{figure}
\begin{figure}[!h]
\centering
\includegraphics[width=1.9in]{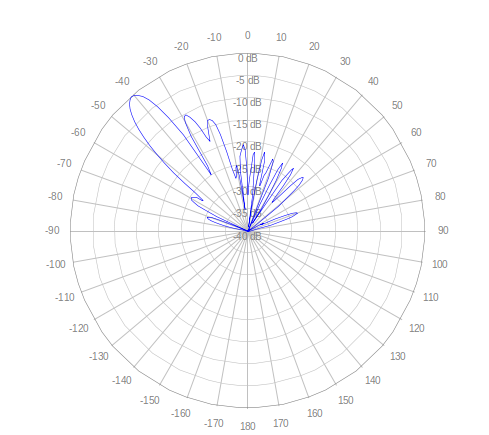}
\caption{Horizontal radiation pattern of scenario \#2, 5G antenna at ``Pos2''.}
\label{fig:beamcambio2}
\end{figure}
Measured throughput values are reported in Table~\ref{table:BS_DL} for DL case whereas in Table~\ref{table:BS_UL} for UL case. 
The static case is identified by a stationary stirrer, labeled as ``OFF'' whereas rotating stirrer denotes the dynamic scenario labeled as ``ON''.
The maximum throughput achievable is $550$~Mbps when the MCS is about $28$, related to a
$64$~QAM current modulation. The throughput is evaluated by the product of two factors,
so resulting into a not integer number:
\begin{itemize}
\item the ratio between DAI received in DL and the DL slot of their transmission;
\item TDD frame, namely the ratio between the number of DL slot and total number of frame slot.
\end{itemize}
Moreover, the MCS value reported in our results is not an integer value because
it is averaged over every acquisition time slot, i.e. every $0.125$~ms.
Both throughput and MCS are close to the maximum achievable. The beam serving distribution, are almost the same, in particular beam~\#3.
For the UL scenario in all the explored cases, reported in Table~\ref{table:UE_DL}, both RSRP and SINR are always good, probably due to the operation of beamforming.

\begin{figure}[h]
\centering
\includegraphics[width=2.2in]{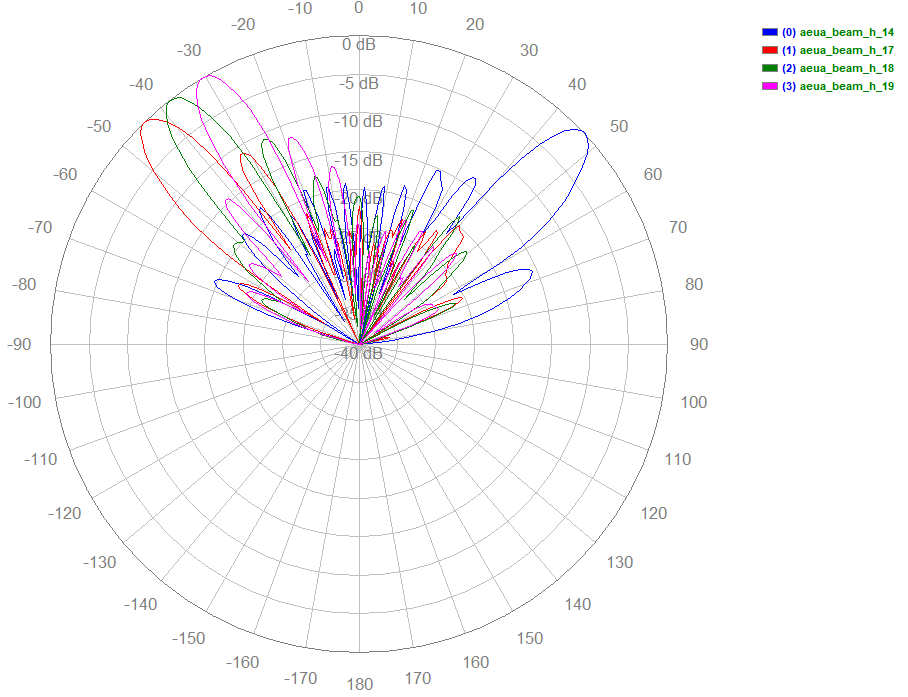}
\caption{Horizontal radiation pattern of the selected beams for scenario~\#2, 5G antenna at ``Pos2'', and the RC door is covered by anechoic panels. In the label are reported the beams activated by the BS.}
\label{fig:beamcambio1}
\end{figure}

In order to check the capability of the antenna beamforming to highlight the best transmission configuration, in ``Pos1'' case we covered the RC door
by absorbing material, see~\figurename~\ref{fig:portacoperta}.
The beam serving distribution changed and the selected beam from the BS is the beam~\#5 instead of beam~\#3,
\figurename~\ref{fig:beamtest} shows the beam switching. This behavior can be explained in that the preferable path is blocked by absorbing materials thus
the BS select a beam that hits again the wall of the chamber, mitigating the effect of the added obstacle and maintaining the same quality of the communication.
No evident changes in terms of RSRP are observed.
This indirectly confirms the capability of the BS antenna to beamform the radiation pattern also inside the RC.

We also investigated the scenario when the RC door, covered by absorbing material, is opened.
In this case we measured:
\begin{itemize}
\item DL: throughput 538.62 Mbps -- MCS of 26.12 -- 15 dB SINR -- rank 2
\item UL: throughput 75.80 Mbps -- MCS of 24.41 -- 15.74 dB SINR -- rank 2.
\end{itemize}

\begin{table}
 \caption{GNB DL performance.}
 \label{table:BS_DL}
  \begin{tabular}{c|c|c|c|c|c}
  Position & Scenario & Stirrer & Throughput (Mbps) & MCS     & Rank     \\ \hline
  1        & \#1      &  ON     &  532.74           & 25.65   &    2     \\ 
  1        & \#1      &  OFF    &  542.31           & 25.73   &    2     \\ 
  1        & \#2      &  ON     &  539.48           & 24.84   &    2     \\ 
  1        & \#2      &  OFF    &  330.88           & 20.71   &    2     \\ 
  2        & \#1      &  ON     &  536.31           & 24.41   &    2     \\ 
  2        & \#1      &  OFF    &  536.8            & 25.68   &    2     \\ 
  2        & \#2      &  ON     &  525.2            & 25.38   &    2     \\ 
  2        & \#2      &  OFF    &  506.65           & 25.24   &    2     \\ 
  \end{tabular}
\end{table}
\begin{table}
 \caption{UE UL performance.}
 \label{table:BS_UL}
  \begin{tabular}{c|c|c|c|c|c}
  Position & Scenario & Stirrer & Throughput (Mbps) & MCS     & Rank    \\ \hline
  1        & \#1      &  ON     &  100.19           & 26.3    &   2     \\ 
  1        & \#1      &  OFF    &  100.25           & 26.98   &   2     \\ 
  1        & \#2      &  ON     &  53.12            & 18.2    &   2     \\ 
  2        & \#2      &  OFF    &  65.41            & 16.5    &   2     \\ 
  \end{tabular}
\end{table}
\begin{table} 
 \caption{UE side: RSRP (dBm) and SNR (dB).}
 \label{table:UE_DL}
  \begin{tabular}{c|c|c|c|c}
  Position & Scenario & Stirrer & RSRP (dBm) &  SNR (dB) \\ \hline
  1        & \#1      &  OFF    &  -55.5     & -6.7      \\ 
  1        & \#2      &  OFF    &  -56.8     & -8.2      \\ 
  1        & \#1      &  ON     &  -55.9     & -6.9      \\ 
  1        & \#2      &  ON     &  -57.2     & -9.0      \\ 
  2        & \#1      &  OFF    &  -57.6     & -7.5      \\ 
  2        & \#2      &  OFF    &  -59.1     & -10.8     \\ 
  2        & \#1      &  ON     &  -58.1     & -8.1      \\ 
  2        & \#2      &  ON     &  -59.8     & -12.7     \\ 
  \end{tabular}
\end{table}

We are able to reduce the power broadcast but not the overall power, so it is not possible to degrade a lot the radio propagation.
Up to $-40$~dBm there are no significant throughput variation.
When the SSB block power is reduced to~$-60$~dBm, differently the SINR measured by the UE (related to the SQI)
decreased as well as the throughput,~$259$~Mbps, and the MCS of~$17.88$ w.r.t. MCS of $26$ in previous case.
No impact on MIMO distribution are observed.

We also investigated the ``Pos2'' in two scenarios:
\begin{itemize}
\item the RC door was free;
\item the RC door was covered with absorbing materials.
\end{itemize}
In the first case the BS activated always the same beam ($-40$~deg) to ensure a good connection, as shown in~\figurename~\ref{fig:beamcambio2}.
In the second case, in presence of absorbing material, the BS is compelled to switch the beam at different angles to ensure the connection,
as shown in the radiation pattern reported in~\figurename~\ref{fig:beamcambio1}.
It seems that the variation of the $B_c$ from ``Scenario \#1'' to ``Scenario \#2'' does not appreciably impact on results due to the larger bandwidth adopted by the BS, i.e.~$100$~MHz.

By comparing the throughput, RSRP and MCS reported in Tables~\ref{table:BS_DL},~\ref{table:BS_UL} and~\ref{table:UE_DL} respectively, 
it is evident that when the horizontal stirrer was rotating, it did not remarkably affect the quality of the
transmission once the BS selected the optimal beam.
This robustness against the stirrer rotation is evident in both DL and UL in combination of ``Pos1'', ``Pos2'', and in both
investigated scenarios, i.e. ``Scenario~\#1'' and ``Scenario~\#2''.
This can be carried to a commercial scenario, e.g a factory, where there are moving conductive objects, e.g. AVGs, that intercept the trajectory of the 5G signal for a while, like our stirrer acts.

\section{Conclusion}
As we are approaching the deployment of 5G systems in the mmWave range, we conducted an experimental analysis in this frequency range by means of
commercial 5G devices such as BS, antenna and UE and thanks to the support from the mobile operator (TIM) and vendor (Nokia).
In particular, this measurements campaign focused on the KPIs evaluation of a commercial 5G BS that operates at the mmWave frequency range, center frequency at $26.95$~GHz.
Measurements are performed in the university facility, i.e. the RC which allows us to emulate a residential and a commercial real-life environment.
We investigated the capability of the BS system to exploit the propagation channel in which it operates.
Main results of these measurements campaign show that RSRP, throughput, MCS and SINR collected by the UE and by BS are good and stable in the explored radio conditions.
In our tests we evaluated the functionalities of the BS which is able to maintain a good communication link in the investigated scenarios.
We showed that the BS scans the area and selects the beam, or more than one, for the transmission and it is able to change the beam in presence of link blockage.
We forced the BS to change the beam in two ways. We can select the wanted beam of the BS antenna by means of a dedicated tool provided by the Nokia software which controls the entire BS functionalities.
In the first way, we covered the RC door with absorbing materials. In fact, in one scenario just one beam was selected by the BS to ensure the connection with the UE but when the setup was altered by adding absorbing material in the region illuminated by the selected beam, the BS changed the selected beam.
In the second way, we stressed the BS by continuously rotating the RC stirrer that acts as a conductive element which may change the boundary conditions of the environment.
The BS is able to reconfigure itself by maintaining a good radio link connection, close to the maximum achievable, in case of
whatever changes may occur in the surrounding environment.
A key factor that explains the capability to maintain a good connection is the high gain of the antenna, about $28$~dBi, and the possibility to swap across $32$ beams.
The MIMO is always used both in UL and DL direction.
In the UL direction the UE does not achieve the maximum power, this aspect needs to be investigated if this behavior
is due to the power control parameters or to thermal protection mechanism. 


\section*{List of Acronyms} 
\begin{table}[h]
\begin{tabular}{ll}
\multicolumn{2}{c}{} \\ 
\textbf{Acronym}     &  \textbf{Full-form} \\ 
      &                                    \\
2D    & Two Dimensional                    \\
4G LTE& Fourth Generation Long Term Evolution \\
5G    & Fifth Generation                   \\
AVG   & Automated Guided Vehicles          \\
$B_c$ & Coherence Bandwidth                \\
BS    & Base Station                       \\
DAI   & Downlink Assignment Index          \\
DL    & Downlink                           \\
eMBB  & Enhanced Mobile Broadband          \\
EC    & Empty Chamber                      \\
EIRP  & Effective Isotropic Radiated Power \\
GNB   & Next Generation Node B             \\
GPS   & Global Positioning System          \\  
IFT   & Inverse Fourier Transform          \\ 
IoT   & Internet of Things                 \\ 
ITU   & International Telecommunication Union \\
KPI   & Key Performance Indicator          \\
mmWave& millimeter-wave                    \\
M2M   & Machine-to-Machine                 \\ 
MCS   & Modulation Coding Scheme           \\ 
MIMO  & Multiple-Input Multiple-Output     \\ 
ML    & Medium Load                        \\ 
OTA   & Over the air Tests                 \\ 
PDP   & Power Delay Profile                \\
$Q$-factor & Quality factor                  \\
QAM   & Quadrature Amplitude Modulation    \\
QXDM  & QUALCOMM eXtensible Diagnostic Monitor \\
RC    & Reverberation Chamber              \\
RF    & Radio Frequency                    \\
RIS   & Reconfigurable Intelligent Surface \\
RSRP  & Reference Signals Received Power   \\
SINR  & Signal to Interference plus Noise Ratio \\ 
SNR   & Signal to Noise Ratio              \\
SSB   & Synchronization Signal Block       \\
SQI   & Signal Quality Indicator           \\
TDD   & Time Division Duplex               \\
UE    & User Equipment                     \\
UL    & Uplink                             \\
UHL   & Ultra High Load Chamber            \\
VNA   & Vector Network Analyzer            \\
\end{tabular}
\end{table}

\section*{Acknowledgment}

This work was supported by the European Union (EU) HORIZON 2020 (H2020) Reconfigurable Intelligent Sustainable Environments for 6G Wireless Networks (RISE-6G) Project
under Grant 101017011.


\begin{IEEEbiography}[{\includegraphics[width=1in,height=1.25in,clip,keepaspectratio]{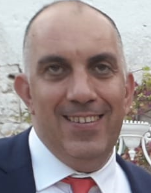}}]{Michele Colombo}
received the M.Sc. degree in Electronic Engineering from the Politecnico di Milano, Milan, Italy, in 1998.
He is currently with the Department of Network Performance and Optimization, Nokia Networks, Rome.
He is involved in radio access design, optimization and testing for UMTS, LTE and 5G technologies.
\end{IEEEbiography}

\begin{IEEEbiography}[{\includegraphics[width=1in,height=1.25in,clip,keepaspectratio]{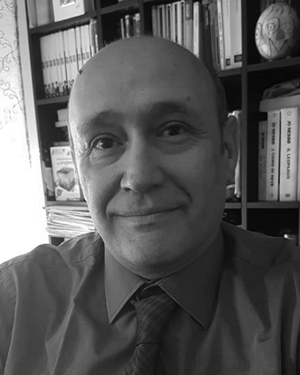}}]{Riccardo Diamanti}
received the degree in Telecommunications
from the University of Ancona (now Universit\`{a}
Politecnica delle Marche), Ancona, Italy,
in 2000, and the degree in Electronics Engineering
from the Universit\`{a} Politecnica delle Marche, Ancona,Italy, in 2015.  
He is currently with Telecom Italia,
as specialist in network performance management.
\end{IEEEbiography}

\begin{IEEEbiography}[{\includegraphics[width=1in,height=1.25in,clip,keepaspectratio]{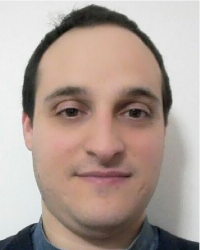}}]{Luca
Bastianelli}
received the M.S. degree in electronic engineering from
Universit\`a Politecnica delle Marche,
Ancona, and the Ph.D. degree in biomedical, electronics and
telecommunication engineering from the same university in 2018.
In 2018 he was a Research Fellow with the Dipartimento di Ingegneria
dell'Informazione, Universit\`a Politecnica delle Marche.
He is currently a Researcher with the CNIT at the Dipartimento di Ingegneria
dell'Informazione, Universit\`a Politecnica delle Marche.
He is involved in the area of electromagnetic compatibility,
telecommunications and computational electrodynamics.
His research interests include reverberation chambers,
wave chaos, propagation in complex systems, ray tracing, time reversal, metasurfaces, 5G and numerical techniques on
high performance computers.
He was a member of the Cost Action IC1407 ACCREDIT.
He was involved in many PRACE projects and he currently lead the FDTDLIME project on reconfigurable intelligent surfaces.
He is currently an active member of the Working Group for the development of the IEEE Std P2718 on near field characterization.
He is currently involved in the H2020 RISE-6G project.

During the Ph.D. degree, he spent seven months in the University of
Nottingham, Nottingham, UK.
During his fellowship he is involved in teaching support activities.
Dr. Bastianelli received the URSI Commission E Young Scientist Award in
2017.
\end{IEEEbiography}


\begin{IEEEbiography}[{\includegraphics[width=1in,height=1.25in,clip,keepaspectratio]{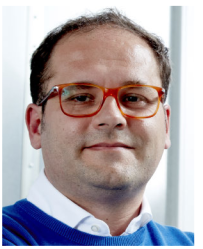}}]{Gabriele Gradoni} received the Ph.D. degree in electromagnetics from Universit\`a Politecnica delle Marche, Ancona, Italy, in 2010. He was a Visiting Researcher with the Time, Quantum, and Electromagnetics Team, National Physical Laboratory, Teddington, UK, in 2008. From 2010 to 2013, he was a Research Associate with the Institute for Research in Electronics and Applied Physics, University of Maryland, College Park, MD, USA. From 2013 to 2016, he was a Research Fellow with the School of Mathematical Sciences, University of Nottingham, UK, where he was a Full Professor of mathematics and electrical engineering. Since May 2023 he has been Full Professor and Chair of Wireless Communications at the 6G Innovation Centre, Institute for Communication Systems, University of Surrey, Guildford, UK. Since June 2020, he has been an Adjunct Associate Professor with the Department of Electrical and Computer Engineering, University of Illinois at Urbana–Champaign, USA. Since 2020, he has been a Royal Society Industry Fellow at British Telecom, UK. Since December 2022, he has been a Visiting Fellow with the Department of Computer Science and Technology, University of Cambridge, UK. His research interests include probabilistic and asymptotic methods for propagation in complex wave systems, metasurface modelling, quantum/wave chaos, and quantum computational electromagnetics, with applications to electro-magnetic compatibility and modern wireless communication systems. He is a member of the IEEE and the Italian Electromagnetics Society. He received the URSI Commission B. Young Scientist Award in 2010 and 2016, the Italian Electromagnetics Society Gaetano Latmiral Prize in 2015, and the Honorable Mention IEEE TEMC Richard B. Schulz Transactions Prize Paper Award in 2020. From 2014 to 2021, he was the URSI Commission E. Early Career Representative.
\end{IEEEbiography}


\begin{IEEEbiography}[{\includegraphics[width=1in,height=1.25in,clip,keepaspectratio]{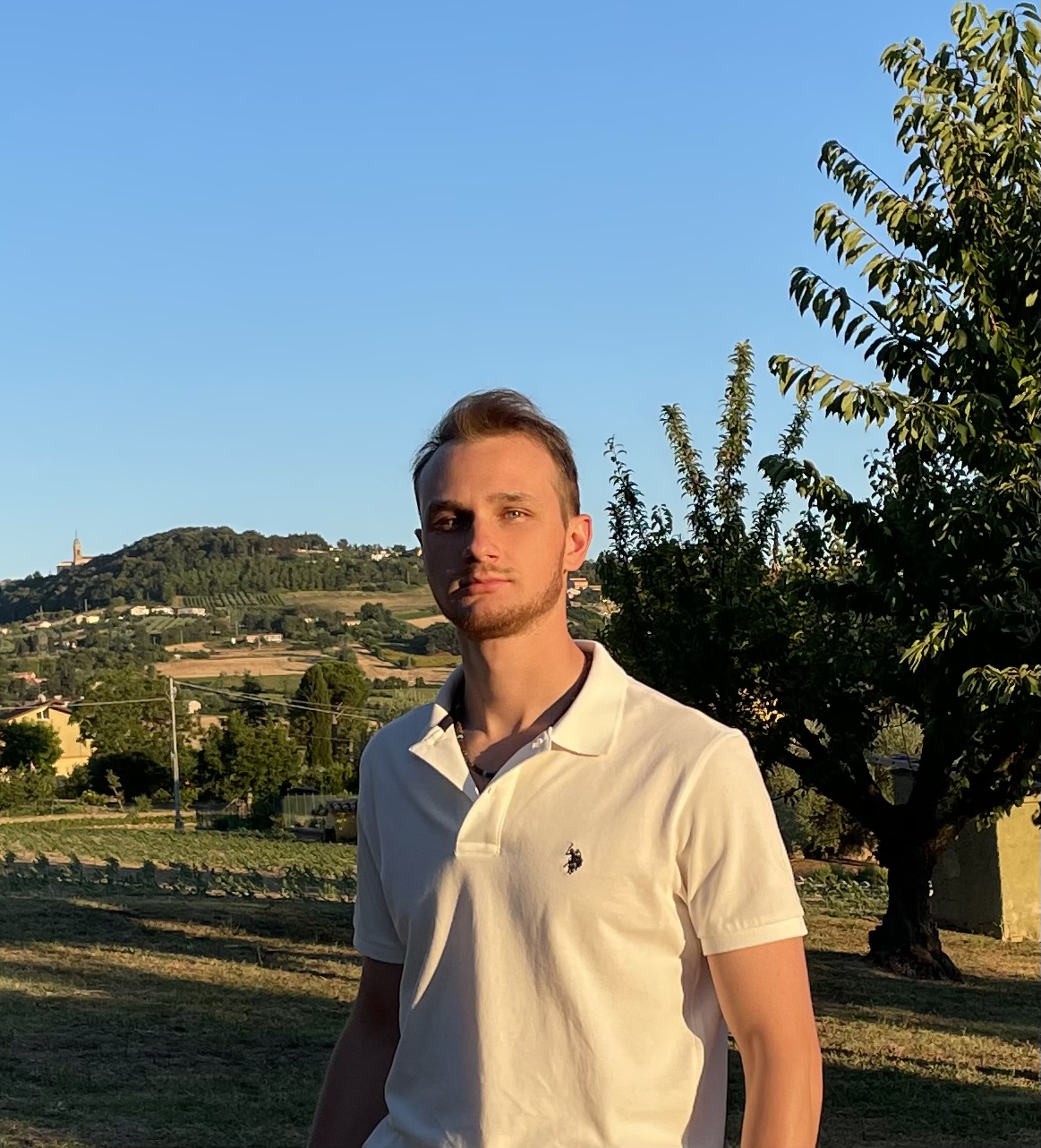}}]{Emanuel Colella} is a Biomedical Engineer graduated in BSc Biomedical Engineering in 2015 at the Università Politecnica delle Marche.
He received the Master's Degree cum laude in Biomedical Engineering in 2019 at the Università Politecnica delle Marche.
In 2020 he started working at the CNIT as researcher for the RISE-6G project.
He worked on computational electrodynamic simulations of RIS to analyze their performance in smart radio environment. In 2022 he has been designated as project lead of working group P2718 on the characterization of unintentional stochastic radiators.
In 2022 he started the PhD in electromagnetics focusing on electromagnetic analysis of complex structures using classical and quantum algorithms. 
\end{IEEEbiography}

\begin{IEEEbiography}[{\includegraphics[width=1in,height=1.25in,clip,keepaspectratio]{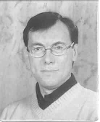}}]{Valter Mariani Primiani}
received the Laurea degree (summa cum laude) in electronic engineering from the University of Ancona, Ancona, Italy, in 1990.

He is currently an Associate Professor in electromagnetic compatibility (EMC) in the Universit\`{a} Politecnica delle Marche, Ancona, Italy.
He is member in the Department of Information Engineering, where he is also responsible for the EMC Laboratory.
His research interests include the prediction of digital printed circuit board radiation,
the radiation from apertures, the electrostatic discharge coupling effects modeling,
and the analysis of emission and immunity test methods.
Since 2003, he has been also involved in research activities on the application of reverberation chambers for compliance testing and for metrology applications.

Prof.\ Mariani Primiani was a Member of the COST Action 1407 ACCREDIT on the characterization of stochastic emissions from digital equipment.
He is senior member of the IEEE (EMC society) and a member of the Italian Society of Electromagnetics.
From 2007 to 2013, he was an active member of the Working Group for the development of the IEEE Std 299.1-2013 on shielding effectiveness measurements.
Since 2021, he was been involved in the RISE--6G project.
Since 2014, he has been a member of the International Steering Committee of EMC Europe.
He is currently an Associate Editor of the journal IET Science, Measurement \& Technology.
\end{IEEEbiography}

\begin{IEEEbiography}[{\includegraphics[width=1in,height=1.25in,clip,keepaspectratio]{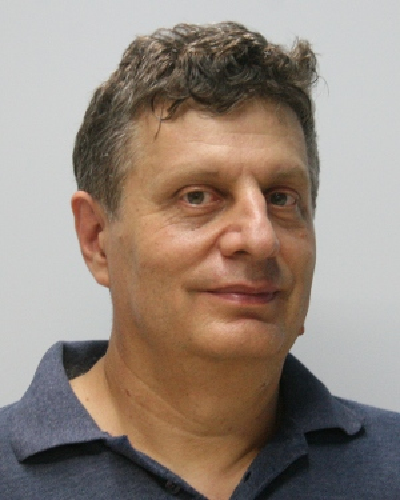}}]{Franco Moglie}
received the ``Dottore Ingegnere'' degree in electronics engineering from the University of Ancona, Ancona, Italy, in 1986,
and the Ph.D. degree in electronics engineering and electromagnetics from the University of Bari, Bari, Italy, in 1992.

Since 1986, he has been a Research Scientist in the Universit\`{a} Politecnica delle Marche, Ancona, Italy, and since 2011,
he has been with the Department of Information Engineering. 
Since 2016, he has been an Associate Professor in the same University.
His current research interests include EM numerical techniques.
In particular, his research activity is in the field of the application of reverberation chambers for compliance testing,
metrology applications, and multipath propagation.

In 2011, Prof.\ Moglie was visiting researcher at Wave Chaos Group, IREAP, University of Maryland, College Park, MD, USA.
From 2013 to 2019, he led three PRACE high performance computing projects on reverberation chambers.
Since 2014, he has been an Italian management committing member of the COST Action IC1407.
From 2007, he is or was an active member of some Working Groups for the development of IEEE Standards: 299.1, 1302, 2715, and 2716.
Since 2017, he has been a secretary of the working group for the IEEE Standard 2718.
Since 2007, he has been a member of the ``Accademia Marchigiana di Scienze, Lettere ed Arti -- Istituto Culturale Europeo''
(Marches Academy of Sciences, Arts and Letters -- European Cultural Institute), which is based in Ancona.
Since 2021, he has been the scientific manager for CNIT at Universit\`{a} Politecnica delle Marche in the RISE--6G project.
He is a member of the IEEE Electromagnetic Compatibility Society and the Italian Electromagnetics Society.
In 2013, he received the title of Distinguished Reviewer of the IEEE Transactions on Electromagnetic Compatibility.
\end{IEEEbiography}


\begin{IEEEbiography}[{\includegraphics[width=1in,height=1.25in,clip,keepaspectratio]{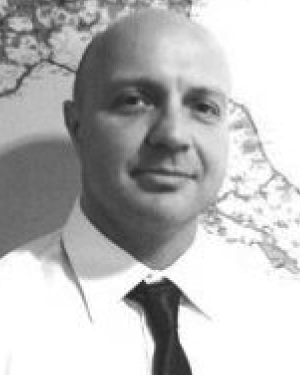}}]{Davide Micheli}
received the ``Dottore Ingegnere'' degree in electronics engineering from the University of Ancona (now Universit\`{a}
Politecnica delle Marche), in 2001,
and the master's degree in astronautic engineering and
the Ph.D. degree in aerospace engineering from the Sapienza University of Rome,
Italy, in 2007 and 2011, respectively.

He is currently with the Department of Wireless Access Engineering, Telecom Italia, Rome, Italy.
He is involved in 2G, 3G, 4G, and 5G radio access network parameterization and optimization.
\end{IEEEbiography}


\end{document}